\documentclass[%
 aip,
 amsmath,amssymb,
 reprint,
]{revtex4-1}

\usepackage{graphicx}
\usepackage{dcolumn}
\usepackage{bm}
\usepackage{enumitem}

\usepackage[
  colorlinks=true,
  urlcolor=blue,
  linkcolor=blue,
  citecolor=blue
]{hyperref}

\usepackage[utf8]{inputenc}
\usepackage[T1]{fontenc}
\usepackage{mathptmx}
\usepackage{etoolbox}
\usepackage{siunitx}
\usepackage{multirow}

\def\baonethirtyeight{$^{138}$Ba$^{+}$}
\def\baonethirtythree{$^{133}$Ba$^{+}$}

\def\barium{Ba$^+$}

\def\bacl{BaCl$_2$}
\newcommand{\appropto}{\mathrel{\vcenter{
  \offinterlineskip\halign{\hfil$##$\cr
    \propto\cr\noalign{\kern2pt}\sim\cr\noalign{\kern-2pt}}}}}

\def\stvbare{$\mathcal{V}^{(\circ)}$}
\def\stvKE{$\mathcal{V}^{(\textrm{KE})}$}
\def\stvPI{$\mathcal{V}^{(\textrm{KE|PI})}$}
\def\stvMM{$\mathcal{V}^{(\textrm{KE|PI|mm})}$}

\def\dfivehalf{$^2D_{5/2}$}

\makeatletter
\def\@email#1#2{%
 \endgroup
 \patchcmd{\titleblock@produce}
  {\frontmatter@RRAPformat}
  {\frontmatter@RRAPformat{\produce@RRAP{*#1\href{mailto:#2}{#2}}}\frontmatter@RRAPformat}
  {}{}
}%
\makeatother

\begin{document}

\preprint{AIP/123-QED}
\title{Ablation loading of barium ions into a surface-electrode trap}

\author{X. Shi}
\affiliation{Center for Ultracold Atoms, Research Laboratory of Electronics, Massachusetts Institute of Technology, Cambridge, Massachusetts 02139, USA}

\author{S.L Todaro}
\altaffiliation{Author to whom correspondence should be addressed: stodaro@mit.edu}
\affiliation{Center for Ultracold Atoms, Research Laboratory of Electronics, Massachusetts Institute of Technology, Cambridge, Massachusetts 02139, USA}

\author{G. L. Mintzer}
\affiliation{Center for Ultracold Atoms, Research Laboratory of Electronics, Massachusetts Institute of Technology, Cambridge, Massachusetts 02139, USA}

\author{C. D. Bruzewicz}
\affiliation{\mbox{Lincoln Laboratory, Massachusetts Institute of Technology, Lexington, Massachusetts 02421, USA}}
\affiliation{Center for Quantum Engineering, Research Laboratory of Electronics, Massachusetts Institute of Technology, Cambridge, Massachusetts 02139, USA}

\author{J. Chiaverini}
\affiliation{\mbox{Lincoln Laboratory, Massachusetts Institute of Technology, Lexington, Massachusetts 02421, USA}}
\affiliation{Center for Quantum Engineering, Research Laboratory of Electronics, Massachusetts Institute of Technology, Cambridge, Massachusetts 02139, USA}

\author{I.L. Chuang}
\affiliation{Center for Ultracold Atoms, Research Laboratory of Electronics, Massachusetts Institute of Technology, Cambridge, Massachusetts 02139, USA}

\date{\today}

\begin{abstract}
Trapped-ion quantum information processing may benefit from qubits encoded in isotopes that are practically available in only small quantities, e.g. due to low natural abundance or radioactivity.  
Laser ablation provides a method of controllably liberating neutral atoms or ions from low-volume targets, but energetic ablation products can be difficult to confine in the small ion-electrode distance, micron-scale, microfabricated traps amenable to high-speed, high-fidelity manipulation of ion arrays.  
Here we investigate ablation-based ion loading into surface-electrode traps of different sizes to test a model describing ion loading probability as a function of effective trap volume and other trap parameters.  
We demonstrate loading of ablated and photoionized barium in two cryogenic surface-electrode traps with $730~\si{\micro\meter}$ and $50~\si{\micro\meter}$ ion-electrode distances.  Our loading success probability agrees with a predictive analytical model, providing insight 
for the confinement of limited-quantity species of interest for quantum computing, simulation, and sensing. 
%
\end{abstract}

\maketitle


Singly ionized barium has recently emerged as a leading ion species for trapped-ion quantum information processing, with a low-lying level structure controllable by visible and near-infrared lasers and a long-lived\ \dfivehalf\ metastable state with a radiative decay lifetime of more than 30~seconds~\cite{Auchter2014}. Additionally, since relevant laser wavelengths for manipulation of Ba$^{+}$ are, in general, longer than those for most comparable ion species, integration of control technologies may be simplified.   Particular attention has been paid to the radioactive isotope \baonethirtythree\, which has nuclear spin $I=1/2$ (refs.~\cite{Hucul2017,Christensen2019,White2022}). Due to its low, non-zero nuclear spin, this isotope has `clock' transitions between hyperfine states that are first-order insensitive to magnetic field fluctuations\cite{Langer2005}\ 
with a minimally complex electronic structure.  
This favorable electronic structure has enabled fast fiducial electronic state preparation with low state preparation and measurement 
error \cite{Christensen2019}. However, \baonethirtythree has a half life of 10.5 years, so safe use suggests employing very small (microgram) quantities to avoid contamination or the requirement for excessive shielding.
Ions of nonradioactive Ba isotopes are often loaded via an oven, from which atoms in a vapor are ionized within the trap region by means of electron-beam bombardment~\cite{DeVoe2002} or photo-ionization~\cite{Steele2007,Wang2011,Leschhorn2012}. Standard ovens are inefficient, however, producing uncollimated atomic beams and generally coating the apparatus in excess source material, and so are not ideal for a radioactive source.

\begin{figure}[tbp!]
\begin{center}
\includegraphics[width=3.3in]{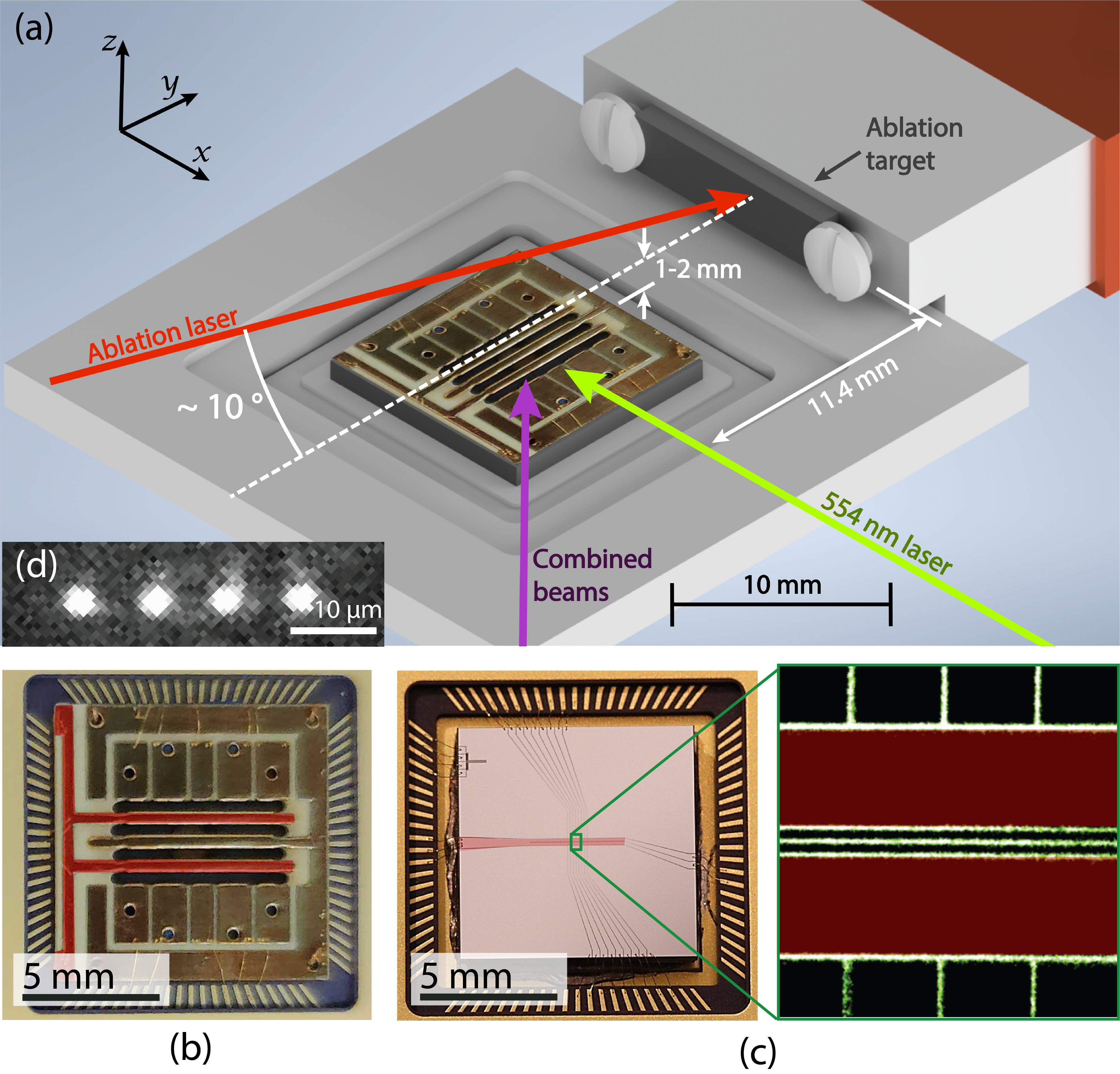} 
\end{center}
\caption{ %
Traps and experimental configuration. (a) Beam orientations and trap mounting hardware. (b, c) The printed-circuit-board and microfabricated surface electrode traps. The rf drive is applied to the electrode shown in red in both systems (false color).
Ions are trapped 730 and 50 $\si{\micro\meter}$, respectively, above the trap surface.  Traps are mounted in ceramic pin grid array (CPGA) mounts for rapid trap exchange. 
(d) Four Ba$^{+}$ ions in the microfabricated trap.
}
\label{fig:apparatus}
\end{figure}

Alternatively, either neutral or ionized atoms can be generated by laser ablation of a source target\cite{Leibrandt2007, Zimmerman2012, Shao2018, Vrijsen2019} which produces a directed plume of material. 
Ablation results in less excess material deposition and introduces a smaller heat load to the apparatus when compared to use of an oven, a benefit for cryogenic or portable system operation.
Initial demonstrations of trapping and coherent control of \baonethirtythree were based on direct ionization via ablation~\cite{Hucul2017,Christensen2019}. 
Photo-ionization of ablated neutral barium atoms and subsequent ion trapping from a small volume target compatible with a radioactive source has also been demonstrated~\cite{White2022}.
These demonstrations of trapping ablated barium were performed in relatively large, three-dimensional radio-frequency (rf) Paul traps~\cite{Christensen2020,White2022}.  Surface electrode traps~\cite{Chiaverini2005} 
are a promising alternative to these bulkier traps for scaling ion-based quantum computers, as they can utilize modern microfabrication processes~\cite{Mehta2014} and introduce the opportunity to miniaturize and integrate technologies such as control electronics~\cite{Stuart2019}, laser light delivery \cite{Niffenegger2020, Mehta2020, Ivory2021}, and single-photon detectors\cite{Todaro2021, Setzer2021, Reens2022} on-chip.
However, surface traps have saddle (escape) points at relatively low potential energies compared to 3D Paul traps of similar size and rf drive parameters~\cite{Chiaverini2005}, making it more challenging to contain highly energetic ions.  Further, rf breakdown in microfabricated traps can limit the ultimate achievable depth~\cite{Wilson2022}. 
Loading low-abundance or radioactive species into surface traps therefore requires specialized techniques.

Here we describe loading of $^{138}$Ba$^{+}$ ions, via laser ablation, into segmented linear surface electrode traps in a cryogenic vacuum system. We characterize the efficiency of ion loading via photoionization of neutral atoms in the ablation plume, for targets constructed though varied techniques, some amenable to radioactive source preparation.  Additionally, using both strontium and barium ions, we analyze ion loading into two surface-electrode-trap designs of very different sizes; the results are used to validate a model proposed to allow prediction of relative loading efficiency as a function of rf-trap parameters.  The described techniques are generally applicable to work toward quantum information processing with registers of rare ionic species, including \baonethirtythree, in microfabricated surface-electrode traps. 

The experiments are conducted in a standard cryogenic ion-trapping system, documented previously~\cite{Sage2012, Chiaverini2014,Bruzewicz2016, Stuart2021}; here we describe aspects which are particular to this work. 
Barium ions are loaded through laser ablation followed by photoionization, and strontium ions are photoionized from a remote, pre-cooled neutral source~\cite{Sage2012, Bruzewicz2016}.  The ablation loading setup is shown in Fig.~\ref{fig:apparatus}(a). We use a Q-switched Nd:YAG pulsed laser (Continuum Minilite II) operated in its fundamental mode at 1064~nm with a pulse duration of 5-7~ns. The beam is focused down to a waist radius of 100~um at the target location with a fluence of 1~$\text{J cm}^{-2}$. The ablation laser propagates along the axial direction of the trap (the $y$ axis as shown on Fig. \ref{fig:apparatus}(a)) and is tilted at an angle in the $yz$ plane to avoid scattering on the trap surface. Neutral barium atoms produced by ablation are subsequently ionized through a two-step photoionization (PI) process~\cite{Leschhorn2012}, using 554~nm and 405~nm lasers, oriented as shown in Fig.~\ref{fig:apparatus}(a).  
The 554~nm laser beam is oriented perpendicularly to the plume generated by ablation to minimize Doppler shifts. All other lasers (for Sr$^+$ photo-ionization, Doppler cooling, and repumping; and for Ba$^+$ Doppler cooling and repumping) are overlapped at the ion location, making a 45 degree angle with respect to the trap axis.
Scattered photons are imaged, using a dual-wavelength high-NA lens (optimized for 422 nm and 493 nm, the wavelengths of the Sr+ and Ba+ detection transitions, respectively) onto either a photo-multiplier tube (PMT) or an electron-multiplying charge-coupled-device (EMCCD) camera.

Several ablation target materials and production methods were investigated (see Appendix~\ref{sec:targets}); a metallic target was used for the loading efficiency measurements presented here.  This target is barium metal (dendritic, 99.8 \% purity), prepared by compressing a single piece to form a rectangular prism, approximately 10 mm by 3 mm by 1 mm in size, mounted 11.4 mm away from the trap center (cf Fig.~\ref{fig:apparatus}a). The metallic target typically obtains a layer of barium oxide during the approximately 5 min of exposure to air required for installation. This oxide layer is then ablated with 5-10 pulses at a high ablation-beam fluence (5-7 $\text{J cm}^{-2}$) while neutral barium fluorescence at 554 nm is monitored using the PMT until sufficient flux is observed. A black spot can usually be seen on the target afterward, and ions are loaded from neutral atoms produced by ablating this spot at the lower (standard) fluence of 1~$\text{J cm}^{-2}$. While an increased loading success rate per pulse can be achieved at higher fluence, we also observe a concomitant increase in the rate of contamination of the trap with particles which can scatter control laser beams or become charged by liberated electrons. We therefore combine low fluence pulses with a higher repetition rate ($10$-$15$~Hz) for rapid loading and minimal contamination.

Two traps were used in this work.  The first is a linear printed circuit board surface-electrode Paul trap (the PCB trap, also known as Bastille~\cite{Leibrandt2007, Splatt_2009, Harlander_2010});
the second is a sputtered-Nb-on-sapphire microfabricated linear Paul trap~\cite{Sage2012, Sedlacek2018} (the microfabricated trap; see Fig~\ref{fig:apparatus}(b,c) for images of both traps).
In the PCB trap, the ion height is $730~\si{\micro\meter}$
and rf drive at a frequency of $7.1~\si{\mega\hertz}$ was used, with \barium\ secular frequencies of 165--295~kHz radial and approximately 100~kHz axial. For the microfabricated trap, the ion height is 50~$\si{\micro\meter}$ and the rf frequency was 41.1~MHz, with \barium\ secular frequencies of 1.9--3.3~MHz radial and 500--600~kHz axial.

\begin{figure*}[htbp!]
\includegraphics[width=7in]{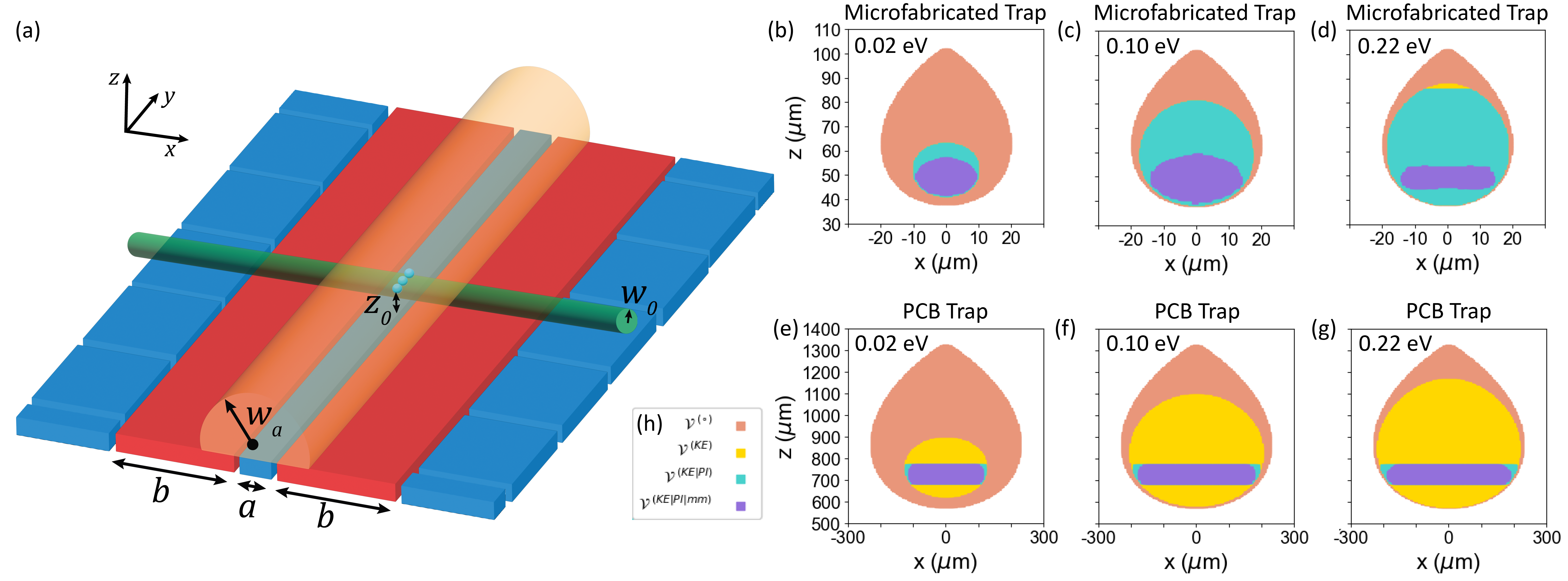}
\caption{Experimental geometry and relevant trap volumes. (a) The trap geometry we consider.  The trap rf null is aligned with the $y$ axis.  The drive rf, with amplitude $V_\textrm{{rf}}$ and angular frequency $\Omega_\textrm{{rf}}$, is applied to the two electrodes shown in red, of width $b$ and separated by a distance $a$. The electrodes shown in blue allow the creation of a DC trapping potential. We assume that neutral atom flux (orange) of uniform density and radius $w_{\textrm{a}}$ is aligned with the $y$ axis.  The photoionization beams (green) with radius $w_0$ are delivered perpendicularly  to the ion flux. 
Ions (not to scale) are trapped a distance $z_0$ above the electrodes.  
(b-d) Cross-sections of trap volumes in the microfabricated trap as a function of increasing $V_{\textrm{rf}}$, from left to right, for an incoming Ba atom moving at $v=150\ \si{\meter\per\second}$. (e-g) Similar cross-sections of trap volumes in the PCB trap.  Trap depths $E_{\textrm{max}}$ are notated for all cross sections (b-g). For all cross-sections, the colors are as indicated in inset (h): the bare trapping volume \stvbare\ is shown in pink, the kinetic energy truncated trapping volume \stvKE\ in yellow, the PI-intersected trapping volume \stvPI\ in teal, and the micromotion-corrected stable trapping volume \stvMM in purple.  Note that these are overlapping volumes.
}
\label{fig:gumdrop}
\end{figure*}

We model the ion loading probability in both traps by first determining the relevant trap volume in which photoionized atoms would nominally be confined indefinitely, neglecting heating effects, but including effects of rf excitation (micromotion).  We however neglect laser cooling since trap dynamics are expected to be much faster than typical Doppler cooling timescales, particularly for ions created near the edge of the effective trap volume.  We calculate the trap potential using an analytic model of the surface-electrode trap geometry\cite{House2008}  (see Appendix \ref{sec:model} for details).  For the trapping potentials we investigate, the potential varies slowly and negligibly in the axial direction within the section of the trapping region defined by the PI beams (cf. Fig.~\ref{fig:gumdrop}(a)), so we 
compare 2D slices of the volumes in the radial $xz$ plane.

As illustrated in Fig. \ref{fig:gumdrop}, the bare stable trapping volume \stvbare is the region for which the potential energy (characterized by the pseudopotential $\phi_{\textrm{rf}}(\vec{x})$ and the residual DC potential $\phi_{\textrm{DC}}(\vec{x})$) of an ion created at a particular location $\vec{x}$ is less than the potential energy at the trap escape point $E_{\textrm{max}}$ (the trap depth).  
To update \stvbare considering ions with non-zero velocity, we assume that an atom of kinetic energy $KE$ is trapped if and only if
$ \phi_{\textrm{DC}}(\vec{x})+\phi_{\textrm{rf}}(\vec{x})+KE\leq E_{\textrm{max}}$;
otherwise, the ion will escape within a few trap cycles. The region defined by this \textit{effective} trap depth $E_{\textrm{max}}-KE$ defines the (velocity dependent) kinetic-energy truncated trapping volume \stvKE. 
The trap volume also depends on the region of sufficient PI beam intensities---ions directly generated by ablation are created outside the conservative trap potential and therefore likely cannot be confined with attainable rates of Doppler cooling. %
We define this photo-ionized trapping volume \stvPI\ as the intersection of \stvKE\ with the PI beams (defined via the $1/e^{2}$ beam radius).

For ions generated far from the trap center, the trap dynamics must also be considered, and \stvPI\ is further reduced by effects due to micromotion. We therefore investigate approximate solutions to the Mathieu equations~\cite{Wineland1998,Berkeland1998} to determine ion trajectories for starting positions $\vec{x}$ within \stvPI.  If the the ion leaves \stvKE\ within a single secular period, we reject the starting point from the rf-dynamics-limited volume \stvMM; the remaining points constitute this volume. 

From this volume determination and known behavior of the ablation and photo-ionization processes, we can calculate the probability of trapping, up to an overall scale factor, by integrating over atom velocity $v$:

\begin{align*}
    P_{\rm trap} \propto \frac{I_2 w_0}{w_{\textrm{a}}^2 k_{\textrm{B}}T}\int_{0}^{v_{\textrm{max}}} \mathcal{V}^{(\textrm{KE|PI|mm})} &\left(\frac{1-e^{-\gamma_1 w_0/2v}}{v}\right)\\
    &\times
e^{-m(v-v_0)^2/2k_{\textrm{B}} T} \, dv \,.
\stepcounter{equation}\tag{\theequation}\label{eq:integral_full}
\end{align*}

\noindent Here $m$ is the atomic mass, $v_{\textrm{max}}\equiv\sqrt{2E_{\textrm{max}}/m}$ is the velocity corresponding to a kinetic energy equal to the trap depth $E_{\textrm{max}}$, $v_0$ is the center-of-mass velocity of the ablation plume, $k_{\textrm{B}}$ is Boltzmann's constant, $T$ is the temperature of the ablation plume, $w_{\textrm{a}}$ is the radius of the ablation plume, $w_0$ is the radius of the focused photo-ionization lasers,  $\gamma_1$ is the saturated first-stage PI linewidth, and $I_2$ is the intensity of the second-stage PI laser.  The derivation of Eq. \ref{eq:integral_full} is discussed in detail in Appendix \ref{sec:model}.

We validate this theoretical model by measuring the relative probability of trapping an ion in both traps as a function of rf drive amplitude.
To compare traps with very different driving conditions (due to differing ion-electrode distances and rf frequencies) we utilize the trap depth.
The trap depth due to rf confinement generally depends on both rf amplitude $V_{\textrm{rf}}$ and drive frequency $\Omega_{\textrm{rf}}$ by
\begin{equation}
E_{\textrm{max}}^{(rf)} = \frac{q^2V_{\textrm{rf}}^2}{\pi^2m\Omega_{\textrm{rf}}^2z_0^2}\kappa(a,b)
\label{eq:trapdepth}
\end{equation}
where $q$ is the ion charge, $m$ is the ion mass, $z_0$ is the ion-electrode distance, and $\kappa(a,b)$ is a dimensionless factor dependent on trap geometry parameters $a$ and $b$ (ref.~\cite{Nizamani_2012} and Appendix \ref{sec:model}). The trap depth is reduced by the application of DC trapping voltages. We distinguish between the ``rf trap depth" (the trap depth in the absence of the DC potential) and the true trap depth, and describe both in Appendix~\ref{sec:model}. We report here loading rates as a function of rf trap depth. 
We extract both the axial trap frequency and the rf amplitude (and thus the trap depth) from secular frequency measurements using standard techniques\cite{Ibaraki2011, Fan2021}.

Loading success rate is defined as the number of ions loaded per  attempt.
A single Ba$^{+}$ loading attempt is made using one pulse of the ablation laser, while a single Sr$^{+}$ attempt is made by unblocking a beam that delivers neutral Sr from the precooled source for $1\ \si{\second}$. Therefore, absolute numbers of Ba$^{+}$ and Sr$^{+}$ loaded should not be compared. Ion presence is confirmed by collecting scattered photons (at 493~nm or 422~nm, respectively) for 10~ms, with the optimal Poissonian discrimination threshold calibrated using a single, Doppler-cooled ion~\cite{Burrell2010}.
For Ba$^{+}$ loading in both traps and Sr$^{+}$ loading in the microfabricated trap, we loaded either zero or one ion per attempt.  In these cases, we calculated the loading success rate by counting trials in which the scattered photon counts exceeded the discrimination threshold.
By contrast, in the PCB trap, we regularly loaded multiple Sr$^{+}$ ions for parameters that led to single ion loading in the other cases. Therefore, we calculated the loading success rate of Sr$^{+}$ in the PCB trap by estimating the number of ions loaded per trial: we divide the total number of photons by the average number of bright photons for a single ion. We eject ions after each trial by turning off the rf drive.

\begin{figure}[tbp]
\includegraphics[width=3.4in]{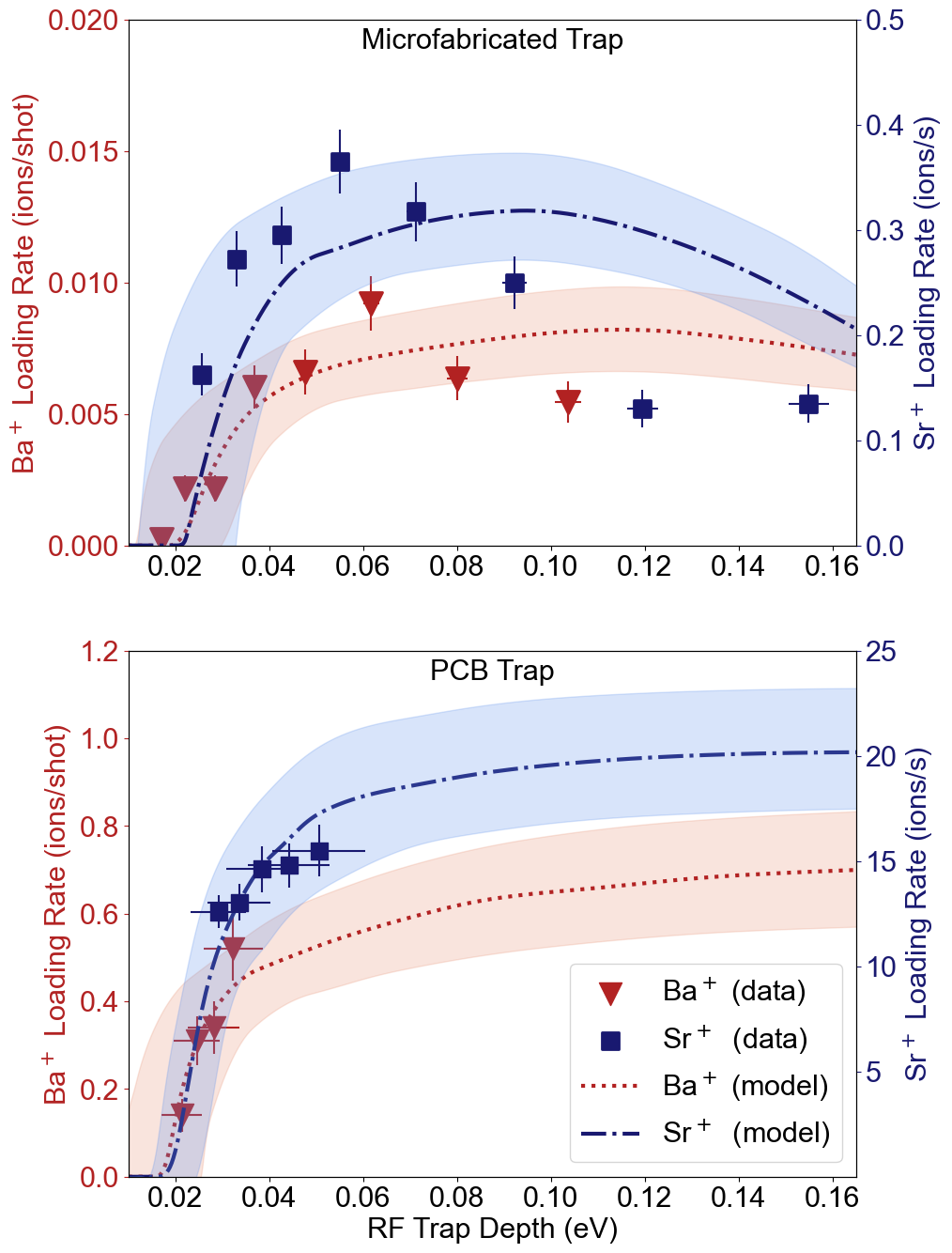}
\caption{Loading rate versus rf trap depth. Measurements for Ba$^+$ (red triangles, left axis) and Sr$^+$ (blue squares, right axis), for both the microfabricated (top) and the PCB (bottom) traps.  Error bars represent one standard deviation.  Dashed lines are single-parameter fits to the model, and the shaded regions incorporate uncertainties in the model due to fluctuations in beam pointing, PI intensity, atomic velocity distributions, shot-to-shot variability in target efficiency, and stray electric field. We smooth discrete jumps in the model values due to finite-step size of the numerical integration using a cubic spline fit.
The analytically calculated threshold rf trap depth is $V_{\rm th}\sim0.02$ eV and the peak loading rate trap depth $E^{(rf)}_{\textrm{opt}}\sim0.08\ {\rm eV}+V_{\rm th}=0.10$ eV for Ba$^+$ in the microfabricated trap, consistent with the data.
}
\label{fig:load_vs_depth_expt}
\end{figure}

Results from these measurements, along with fits to the model (the single-fit parameter being an overall scale factor), are shown in Fig. \ref{fig:load_vs_depth_expt}. 
 Both statistical error in the measured loading rates and estimated errors in the model due to parameter uncertainties are shown; the latter is discussed in detail in Appendix \ref{sec:error}.  At low rf trap depths, the probability of trapping sharply increases with increasing trap depth after a threshold rf amplitude $V_{\rm th}$, determined by total potential positivity (See Appendix~\ref{sec:model}, Eq.~\ref{eq:vthresh}).  At higher trap depths, loading efficiency in the microfabricated trap drops with increasing trap depth.
This
behavior is due to reduction of the stable trapping volume \stvMM at high amplitudes arising from Paul trap instability, as previously described.  
This leads to optimal trapping conditions at a trap depth of
\begin{equation}
E_{\textrm{opt}}\approx \frac{1}{2}m\left(v_0+\sqrt{\frac{k_{\textrm{B}} T}{m}}\right)^2,
\end{equation}
which is approximately 0.08 eV for ablated barium, as detailed in Appendix \ref{sec:analytic}. 
The experimentally observed reduction is qualitatively consistent with predictions from the theoretical model, as is the observed lack of a similar reduction in the PCB trap over experimentally achievable trap depths.
At the upper end of attainable trap depths in the microfabricated trap, the micromotion amplitude is large throughout a substantial fraction of \stvPI, reducing \stvMM.  In contrast, the micromotion amplitude remains small over a similar fraction of \stvPI\ in the physically larger PCB trap for all experimentally achievable depths.
We observe a somewhat faster reduction in loading rate at high trap depths in the microfabricated trap than is predicted, suggesting that there is some behavior in this regime that is not fully accounted for in the model.

We observe a higher loading success rate for the larger PCB trap than for the microfabricated trap at all trap depths, despite
 the higher trap depth attainable in the significantly smaller microfabricated trap.
This is attributable to the larger trapping volume in the PCB trap, in which \stvPI is approximately a factor of 150 larger, owing to its physically larger dimensions.

For this demonstration, we loaded ions from an ablated barium metal source.  This is incompatible with trapping the isotope \baonethirtythree, as radiaoctively enhanced pure Ba metal is not readily available, to the best of our knowledge.
Ablation of BaCl$_2$, a viable source for \baonethirtythree atoms, was previously shown to produce neutral Ba with an average temperature of $T = 37000\ \si{\kelvin}$,\cite{White2022} more than an order of magnitude higher than the melting point of Ba metal.    With liberated atoms at this higher temperature, our model predicts a reduction by roughly an order of magnitude in loading probability; this presents a challenge when loading microfabricated traps from low-volume radioactive sources.  On the other hand, we anticipate, based on the results presented here, that an in-situ chemical reaction in which a more reactive metal is allowed to bond with the chlorine, leaving free barium, could dramatically improve the probability of loading Ba$^+$ from ablation of a BaCl$_2$ source (after such preparation).

The model also has implications for design of future trapped-ion systems, as there is interest in continued miniaturization of microfabrication-compatible traps to address speed and scaling challenges in quantum information processing. 
The relative loading rates observed between the PCB and microfabricated traps, as well as the model describing their performance, confirm that loading rates drop as trap volume is reduced, highlighting the difficulties of loading rare species into increasingly miniaturized traps, regardless of trap depth.
We suggest the consideration of the stable trapping volume as described here as an additional valuable metric in trap design.  For instance, small loading rates in low-volume traps could be compensated through the inclusion of a high-volume loading zone and transition region.

\section*{Author Contributions}
\noindent The first two authors contributed equally to this work.

\noindent
\textbf{Xiaoyang Shi:} Conceptualization, formal analysis, investigation, methodology, software, validation, visualization, writing/original draft preparation, writing/review \& editing;
\textbf{Susanna L. Todaro:} Conceptualization, formal analysis, investigation, methodology, software, validation, visualization, writing/original draft preparation, writing/review \& editing;
\textbf{Gabriel Mintzer:} Software, writing/review \& editing;
\textbf{Colin D. Bruzewicz:} Methodology, supervision, writing/review \& editing;
\textbf{John Chiaverini:} Conceptualization, funding acquisition, methodology,  project administration, supervision, writing/review \& editing;
\textbf{Isaac L. Chuang:} Conceptualization, funding acquisition, methodology, project administration, supervision, writing/review \& editing

\begin{acknowledgments}
This research was supported by the U.S. Army Research Office through grant W911NF-20-1-0037 and by the NSF Center for Ultracold Atoms. S.L.T. is supported by an appointment to the Intelligence Community Postdoctoral Research Fellowship Program at MIT, administered by Oak Ridge Institute for Science and Education through an interagency agreement between the U.S. Department of Energy and the Office of the Director of National Intelligence.  The authors thank Eric Hudson and Wes Campbell and their research groups for helpful discussions.
\end{acknowledgments}

\section*{Data Availability Statement}

\noindent Data available from the authors upon reasonable request.

\bibliography{ba-loading}

\appendix

\section{\label{sec:targets}Target material and characterization}

Barium metal is highly reactive and oxidizes quickly in air.  We prepare our barium targets under an inert atmosphere and transfer the target to the experiment vacuum system with less than five minutes of air exposure.  In this time, a layer of grayish-white barium oxide develops on the surface.  As discussed in the main text, we break through this layer with a series of high-energy preparation ablation pulses.  However, this process can result in deposition of particles onto the trap; these particles can produce visible scatter of the ion cooling and manipulation lasers. Further, this material can introduce stray electric fields, and thus the ability to compensate the total stray field can degrade over time. 
To avoid oxidization, some groups have used barium in a chemically bound form.  Neutral barium was freed in-situ by reacting BaCO$_3$ with tantalum at an elevated temperature~\cite{DeVoe2002}.
More recently, laser ablation of \bacl~ has been demonstrated by multiple groups\cite{Hucul2017, Christensen2019, White2022}.  \bacl\ is used in medical testing, so it is relatively readily available in a radioactively enhanced format.

We also attempted to load \baonethirtyeight{} with several targets made from barium in a bound form.  These attempts included BaCl$_2$ and BaTiO$_3$ compressed powder, a commercial BaTiO$_3$ crystal substrate, a Ba$_{0.5}\text{Sr}_{0.5}\text{TiO}_3$ sputtering target, and a home-grown BaCl$_2$ crystal. For all of these targets except the home-grown \bacl\ crystal, we observed neutral Ba production by laser ablation using fluorescence spectroscopy on the $^1$S$_0\leftrightarrow^1$P$_1$ transition at $554\ \si{\nano\meter}$.  However, these targets all required a higher ablation laser fluence than the Ba metal target did.  We were able to load \barium into the PCB trap from both the BaCl$_2$ and BaTiO$_3$ powder targets and into both traps using the BaTiO$_{3}$ crystal substrate. However, the trap was quickly contaminated by white particles after ablating the powder targets. With all BaTiO$_{3}$ targets, a black discoloration formed at the ablation spot and quickly reduced ablation efficiency.

\begin{figure}[tb]
\includegraphics[width=3.4in]{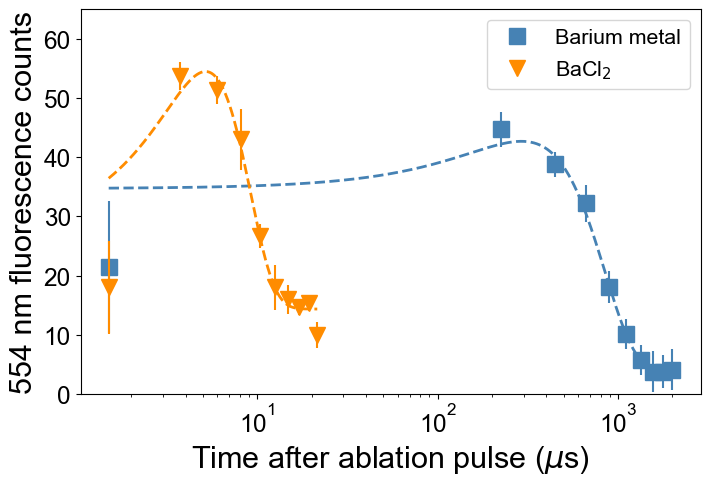}
\caption{Time of flight spectroscopy of ablated barium metal (blue squares) and \bacl\ (orange circles). Fits to Gaussian arrival times are included (dashed lines) to guide the eye. We collect photons for $1.5\ \si{\micro\second}$, so dynamics on shorter timescales cannot be distinguished from the background.  A neutral barium atom with kinetic energy 0.1 eV corresponds to an arrival time of $82\ \si{\micro\second}$.   We do not expect that these measurements correspond linearly to absolute numbers of produced Ba atoms; slower-moving atoms scatter multiple photons while passing through the laser, and appear over-represented in these measurements.  Nonetheless, it is clear that typical atom velocities from the compound target are two orders of magnitude larger than those from the metal target.}
\label{fig:timeofflight}
\end{figure}

No bound barium sources we worked with have been as long-lived and as reliable as the metallic barium source.  We attribute this difference in behavior to differing thermal distributions of barium ablated from barium metal versus barium ablated from \bacl\ or BaTiO$_3$.  Time-of-flight spectroscopy measurements comparing a compound target to a metal target are shown in Fig. \ref{fig:timeofflight}; fluorescence from Ba atoms at 554~nm is measured as a function of delay after an ablation pulse. We observe significantly larger energies for the atoms from the compound. While our model predicts that we should be able to trap some fraction of even the very energetic atoms produced by ablation of \bacl, the predicted loading probability drops by an order of magnitude or more, increasing sensitivity to experimental imperfections and calibration fluctuations.  Further, the rate of stray charge production at high fluence makes practical use of these sources challenging, since it requires frequent recalibration of stray electric fields.  We anticipate that the best prospect for a robust \barium\ ablation source using \bacl\ in a surface electrode trap will require in-situ production of free Ba.

\section{\label{sec:model}Theoretical Model}

\begin{figure*}[t!]
    \centering
    \includegraphics[width=7in]{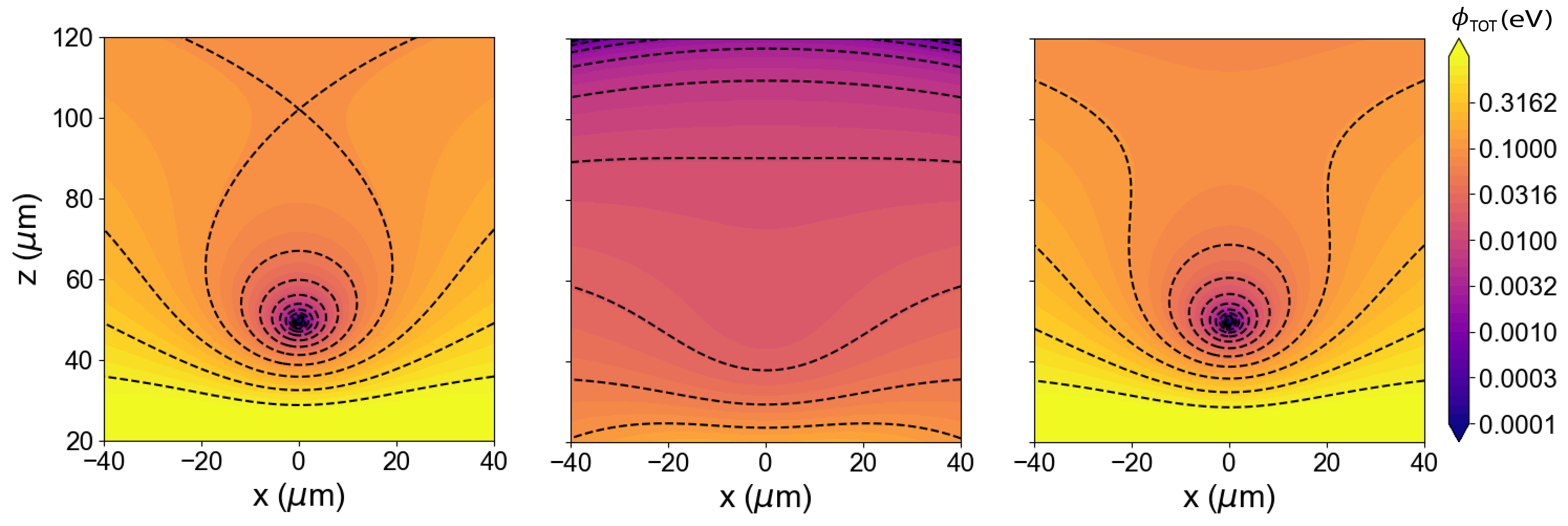}
    \caption{Slice of the microfabricated trap total effective potential for \barium\ in the $xz$ plane at $y=0$ with $\Omega_{\textrm{rf}}=40\ \si{\mega\hertz}$ (a) with $V_{\textrm{rf}}=100\ \si{\volt}$ in the absence of a DC axial trapping potential, (b) with $V_{\textrm{rf}}=20\ \si{\volt}$ and a $600\ \si{\kilo\hertz}$ axial trapping potential ($V_{\textrm{rf}}<V_{\textrm{th}}$) and (c) with $V_{\textrm{rf}}=100\ \si{\volt}$ and a $600\ \si{\kilo\hertz}$ axial trapping potential ($V_{\textrm{rf}}>V_{\textrm{th}}$). The trap surface is located at $z=0$. The trap center is located at $(x, z)=(0, 50)\ \si{\micro\meter}$ and the escape point is located at $(x,z)=(0, 102)\ \si{\micro\meter}$. 
    }
    \label{fig:ion_potential_isosurface}
\end{figure*}

We limit this discussion to the 5-wire symmetric surface-electrode trap~\cite{Chiaverini2005}, in which rf voltage is applied to two electrodes of width $b$, with the closest edges of the two electrodes a distance $a$ apart (red in Fig. \ref{fig:gumdrop}(a)).  This generates a translationally symmetric trapping pseudopotential $\phi_{\textrm{rf}}$ above the trap surface, which we estimate according to an analytic model for surface electrode traps~\cite{House2008}.  This model treats all electrodes as gap-free regions in a grounded plane, with the rf electrodes infinite in axial extent.   The rf pseudopotential for the microfabricated trap under this model is as shown in Fig. \ref{fig:ion_potential_isosurface}(a). 
In keeping with typical notation, we refer to the line of translational symmetry as the ``trap axis" and label motion along (perpendicular to) this axis ``axial" (``radial"). In our geometry, the trap axis is aligned with the $y$ axis, and $x$ and $z$ are the radial directions, as shown in Fig. \ref{fig:gumdrop}.
We apply DC voltages to the segmented electrodes (blue in Fig. \ref{fig:gumdrop}) in order to confine the ion along the trap axis via creation of a potential $\phi_{\textrm{DC}}$.
The DC voltages, in addition to providing axial confinement, introduce a negative curvature which reduces confinement in at least one of the radial directions (cf. Laplace's equation).  Within the extent of the photo-ionization beams, the axial potential is well-approximated as harmonic, and
\begin{equation}
\left(\frac{\partial^2\phi_{\textrm{DC}}}{\partial x^2}\right) =
\left(\frac{\partial^2\phi_{\textrm{DC}}}{\partial z^2}\right) = 
-\frac{1}{2}m\omega_{\textrm{ax}}^2
\end{equation}
and the full radial potential at position $(x,z)$ is then $\phi_{\textrm{TOT}}=\phi_{\textrm{rf}}(x,z)-m\omega_{\textrm{ax}}^2(x^2+z^2)/4$.   (We have limited ourselves in this work to the symmetric case in which the curvatures due to the DC potential in $x$ and $z$ are equal, but note that in general, the ratio between these two curvatures can be chosen by appropriate design of the DC trapping potentials.)

The pseudopotential has a minimum at a distance
\begin{equation}
z_0=\frac{1}{2}\sqrt{a(a+2b)}
\end{equation}
from the trap surface, providing confinement in the $xz$ plane, and a saddle point at
\begin{equation}
 z_{\textrm{esc}} = \sqrt{z_0(a+b+z_0)},
\end{equation}
limiting the region in which ions can be trapped.  The rf trap depth is then the value of the pseudopotential at this saddle point (eq. \ref{eq:trapdepth}),
with $\kappa(a,b)$ a dimensionless trap-dependent geometric factor\cite{Nizamani_2012}
\begin{equation}
\kappa(a,b)\equiv\frac{ab^2(a+2b)}{4(a+b)^2(a+b+\sqrt{a(a+2b)})^2}.
\end{equation}

\noindent The full potential at the escape point (the ``true" trap depth) is then
\begin{equation}
E_{\textrm{max}} = \frac{q^2V_{\textrm{rf}}^2}{\pi^2m\Omega_{\textrm{rf}}^2z_0^2}\kappa(a,b) -\frac{1}{4}m\omega_{\textrm{ax}}^2z_{\textrm{esc}}^2.
\end{equation}
We can determine a minimum rf amplitude necessary for trapping by requiring that this full potential be positive, which is fulfilled at rf amplitudes greater than or equal to
\begin{equation}
V_{\textrm{th}}=\frac{\pi m\Omega_{\textrm{rf}}\omega_{\textrm{ax}}z_0z_{\textrm{esc}}}{2q\sqrt{\kappa(a,b)}}.
\label{eq:vthresh}
\end{equation}
Cross-sections of the full potential $\phi_{\textrm{TOT}}$ for drive amplitudes below and above $V_{\textrm{th}}$ are shown in Fig. \ref{fig:ion_potential_isosurface}(b) and \ref{fig:ion_potential_isosurface}(c), respectively.

From this estimate of the trap pseudopotential, we define the four trapping volumes used to estimate the probability of trapping.

{\bf Bare stable trapping volume \stvbare}: 
The first volume \stvbare\ can be determined from the pseudopotential.
This volume is the area contained by the largest isopotential below the trap depth $E_{\textrm{max}}$.
We determine this by a numerical breadth-first search from the trap center $\vec{x}_0$ to the isopotential curve defined by the set of all positions $\vec{x}$ where the pseudopotential is equal to $E_{\textrm{max}}$:  $\{\vec{x}|\phi_{\textrm{TOT}}(\vec{x})=E_{\textrm{max}}\}$.

{\bf Kinetic energy truncated trapping volume \stvKE}: We next introduce the effect of the kinetic energy of the incoming atom. 
We assume that an ion of kinetic energy $KE$ at position $\vec{x}$ will be trapped if and only if $\phi_{\textrm{TOT}}(\vec{x})+KE\leq E_{\textrm{max}}$;
otherwise, the ion will immediately escape within a few trap cycles. (This is consistent with the assumption that all cooling  processes, including laser and sympathetic cooling, are slow compared to trapping dynamics.  We expect this is an appropriate approximation, since linewidths for Doppler cooling are typically on the order of tens of megahertz, meaning there are at most a few scattering events per rf cycle.)  We compute this probability numerically by modifying the calculation of the trapping volume: instead of integrating to the trap depth, we integrate the volume out from $\vec{x}_0$ to the isopotential curve defined by an \textit{effective} trap depth: $\{\vec{x}|\phi_{\textrm{TOT}}(\vec{x}) = E_{\textrm{max}}-KE\}$.  This reduces the stable trapping volume from \stvbare\ to the kinetic-energy dependent trapping volume \stvKE. 

{\bf PI intersected trapping volume \stvPI}:
The stable trapping volume is also dependent on overlap with the photoionization beams. We define the photo-ionized trapping volume \stvPI, which is the intersection of the kinetic-energy truncated trapping volume \stvKE with the PI beams.  We assume that the PI beams are centered on the rf null with beam waists $w_0$, as shown in Fig.~\ref{fig:gumdrop}. In the case of the PCB trap, whose dimensions are large compared to our focused beam waist of $50\ \si{\micro\meter}$, this significantly reduces the trapping volume.  In the case of the microfabricated trap, there is a more modest but non-negligible reduction in trap volume due to the PI beams.

{\bf Micromotion cutoff trapping volume \stvMM}:  For ions generated far from the trap center, the dynamics of the trap must be considered, and the stable trapping volume will be further reduced by effects due to ion micromotion at the rf drive frequency.
In particular, for some fraction of ions far from the rf null, the micromotion amplitude is large enough that at points in the ion's micromotion cycle, it leaves the trapping volume and is lost.  This serves to  reduce the stable trapping volume at large rf amplitudes.

The impact of micromotion on the stable trapping volume is difficult to compute analytically for the surface electrode trap, but we can evaluate the effect numerically. We estimate the trajectories of ions by numerically examining the solutions to the Mathieu equations\cite{Wineland1998, House2008}
\begin{equation}
\frac{\partial^2\vec{x}}{\partial t^2}+\frac{1}{4}\Omega_{\textrm{rf}}^2\left[A+2Q\cos\left(\Omega_{\textrm{rf}}t\right)\right]\cdot\vec{x}=0
\end{equation}
where
\begin{equation}
    Q_{ij} \equiv \frac{2q}{m\Omega_{\textrm{rf}}^2}\left(\frac{\partial^2\phi_{\textrm{rf}}}{\partial x_i\partial x_j}\right)
    \label{eq:matheiuq}
\end{equation}
and
\begin{equation}
    A_{ij} \equiv \frac{4q}{m\Omega_{\textrm{rf}}^2}\left(\frac{\partial^2\phi_{\textrm{DC}}}{\partial x_i\partial x_j}\right)
    \label{eq:matheiua}
\end{equation}
where $x_i$, $x_j$ are the relevant coordinates.  Near the rf null, the diagonal terms $Q_{ii}$ and $A_{ii}$ are the usual Mathieu parameters $q_i$ and $a_i$, respectively.

 Very close to the rf null, the trapping potential is well-approximated as harmonic and $Q_{ii}$ does not vary spatially.  In this limit, the ion exhibits secular oscillations at constant frequency $f_{i}=\omega_{i}/2\pi=(\Omega_{\textrm{rf}}/4\sqrt{2}\pi)\sqrt{q_{i}^2+a_{i}}$ modulated by micromotion at frequency $\Omega_{\textrm{rf}}/2\pi$ with constant amplitude. We  approximately identify regions of parameter space where the usual solutions break down by noting that the position of the ion as a function of time $t$ is given, for small $Q$ and where we have assumed $A$ is negligible,
\begin{equation}
x_i(t)\approx\sin\left(\omega_{i}t+\delta_i\right)\left[\vec{x}_0+\frac{1}{2}((Q\cdot\vec{x}_i)\cdot\hat{x}_i)\cos\Omega_{\textrm{rf}}t\right],
\label{eq:trajectories}
\end{equation}
where $\delta_i$ is a phase factor dependent on initial conditions\cite{Berkeland1998,Wineland1998}.  At intermediate displacements, we calculate a numerical trajectory according to this expression and the local curvature of the electric potential. We then identify a region of ``normal" motion (where the ion's motion is bounded and we characterize the approximations made as valid) as follows.
We consider only trajectories with initial position $\vec{x}_{\textrm{d}}$ inside \stvPI initially at rest.  We then numerically calculate forward the motion a single nominal secular period $1/f$, where $f$ is calculated according to the curvature matrix $Q$ at the initial position.  If, at any point during this trajectory, it leaves the kinetic-energy truncated trapping volume \stvKE, we consider the ion lost and the initial position unstable.  Otherwise, we add the initial position 
to the micromotion-corrected trapping volume \stvMM. We also reject starting points whose trajectories explore regions in which $|Q\cdot\hat{x}_{\textrm{d}}|>0.5$, as we do not expect these ions to follow closed orbits.
We emphasize that we are \textit{not} using this method to fully identify the character of the motion in the rejected region, as it cannot be described within the approximation in Eq. \ref{eq:trajectories}; we are simply estimating a region of approximation breakdown.

Once we have computed the stable trapping volume, we can calculate a probability of trapping and predict its dependence on trap drive parameters. 
We calculate this through Bayes Theorem for a continuous variable as
\begin{equation}
    P_{\rm trap} = \int_{\vec{x}, v} \mathcal{T}(\vec{x},v) \mathcal{A}(\vec{x}, v)\, d^3\vec{x} \, dv \,,
\label{eq:integral_gen}
\end{equation}
where $\mathcal{T}(\vec{x},v)$
is the probability distribution of photo-ionizing and trapping an atom found at position $\vec{x}$ moving with velocity v and $\mathcal{A}(\vec{x}, v)$ is the underlying probability distribution that a neutral atom will be found at position $\vec{x}$ moving with velocity $v$ after an ablation pulse.
Note that the trap factor $\mathcal{T}(\vec{x}, \vec{v})$ depends only on the PI laser beams and the trap geometry and parameters, while the ablation factor $\mathcal{A}(\vec{x}, \vec{v})$ depends only on the ablation laser and the atom source.  The ablation and trap factors depend on parameter values presented in Table \ref{tab:parameters_model}. 

We define the ablation factor by modeling the source as a simplified thermal distribution of ablated neutral atoms with uniform spatial density
\begin{equation}
    \mathcal{A}(\vec{x},v) \propto  \begin{cases}
    (1/w_{\textrm{a}}^2k_{\textrm{B}}T)
e^{-m(v-v_0)^2/(2k_{\textrm{B}} T)} & |\vec{x}| < w_{\textrm{a}} \\
    0 & {\rm otherwise}
    \end{cases}
    \label{eq:ablationmodel1}
\end{equation}
 where the prefactor is determined by normalization. (Experimental data supports our assumption that ablated ions are thermally distributed \cite{Sheridan2011, Vrijsen2019}.)

Given that a neutral atom is produced with velocity $v$ within the stable trapping volume, the probability that it is photo-ionized is equal to the probability that it is excited by the narrow first-stage PI laser multiplied by the probability that an electron in the excited state is then excited by the second-stage PI laser.  We operate the first-stage PI laser (461~nm for Sr; 554~nm for Ba) in a highly saturated regime, while the second stage is far from saturated.  Therefore the probability that the electron is found in the intermediate state can be determined from the Einstein rate equations, while the probability that the atom is then photo-ionized is linear with the time the atom spends passing through the PI beams.  For an ion moving with velocity $v$ moving perpendicularly through a laser of waist $w_0$, this duration is $w_0/v$. 
Thus we define the trap factor by
\begin{equation}
    \mathcal{T}(\vec{x}, v) \propto \begin{cases}
    (I_2 w_0/v)(1-e^{-\gamma_1 w_0/2v})& \vec{x}\in \mathcal{V}^{(\textrm{KE|PI|mm})} \\
    0 & {\rm otherwise}
    \end{cases}
    \label{eq:trapmodel2}
\end{equation}
and then substitute Eqs. \ref{eq:ablationmodel1} and \ref{eq:trapmodel2} into the general expression of Bayes' Law in Eq. \ref{eq:integral_gen}.  We numerically integrate over the micromotion-corrected stable trapping volume and up to a maximum velocity $v_{\textrm{max}}\equiv\sqrt{2E_{\textrm{max}}/m}$
 to extract a predicted probability of trapping:
\begin{align*}
    P_{\rm trap} \propto \frac{I_2 w_0}{w_{\textrm{a}}^2 k_{\textrm{B}}T}\int_{0}^{v_{\textrm{max}}} \mathcal{V}^{(\textrm{KE|PI|mm})} &\left(\frac{1-e^{-\gamma_1 w_0/2v}}{v}\right)\\
    &\times
e^{-m(v-v_0)^2/2k_{\textrm{B}} T} \, dv \,.
\stepcounter{equation}\tag{\theequation}\label{eq:integral_full_appendix}
\end{align*}

\section{Optimal Loading Conditions}
\label{sec:analytic}

Under appropriate assumptions for the loading plume temperature distribution, we can analytically estimate an optimal trap depth to maximize the probability of loading.  We can consider two general trap geometry conditions: one in which the PI beams are large compared to the trapping region (as in the case of the microfabricated trap) and one in which the PI beams are small compared to the trapping region (as in the case of the PCB trap).  For the small trap, 
approximating the trap as harmonic within relevant trapping regions, the kinetic-energy truncated trap volume goes approximately as
\begin{equation}
\mathcal{V}^{(\textrm{KE})}\approx\mathcal{V}^{(\circ)}\left(\frac{v_{\textrm{max}}^2-v^2}{v_{\textrm{max}}^2}\right)
\end{equation}
and the micromotion-corrected trapping volume goes approximately as
\begin{equation}
\mathcal{V}^{(\textrm{mm|PI|KE})}\approx\mathcal{V}^{(\textrm{KE})}\frac{1}{(1+Q(\vec{x}_0))^2},
\end{equation}
since in the harmonic limit, $Q_{ii}$ is not spatially varying and is equal to its on axis value $Q(\vec{x}_0)$.
Then the overall probability of trapping goes approximately as
\begin{equation}
P_{\textrm{trap}} \appropto
\frac{\mathcal{V}^{(\circ)}}{\left(1+Q(\vec{x}_0)\right)^2}
\int_{0}^{v_{\textrm{max}}}\left(1-\frac{v^2}{v_{\textrm{max}}^2}\right)\frac{e^{-m(v-v_0)^2/2k_{\textrm{B}}T}}{v}dv.
\end{equation}
Assuming that the ablation plume is hot compared to the kinetic energy of any trappable atom, i.e. $m(v_{\textrm{max}}-v_0)^2/2 \ll k_{\textrm{B}}T$, then
\begin{equation}
P_{\textrm{trap}} \appropto 
\frac{\mathcal{V}^{(\circ)}}{\left(1+Q(\vec{x}_0)\right)^2}
\int_{0}^{v_{\textrm{max}}}
\left(\frac{1}{v_0}-\frac{v}{v_{\textrm{max}}^2}\right)
{e^{-m(v-v_0)^2/2k_{\textrm{B}}T}}   dv 
\end{equation}
within the region of integration, because $v\approx v_0$ are the only values of $v$ which contribute to the integral over the Gaussian.
After evaluating this integral, the probability goes as
\begin{figure}[tbp!]
    \centering
    \includegraphics[width=3.4in]{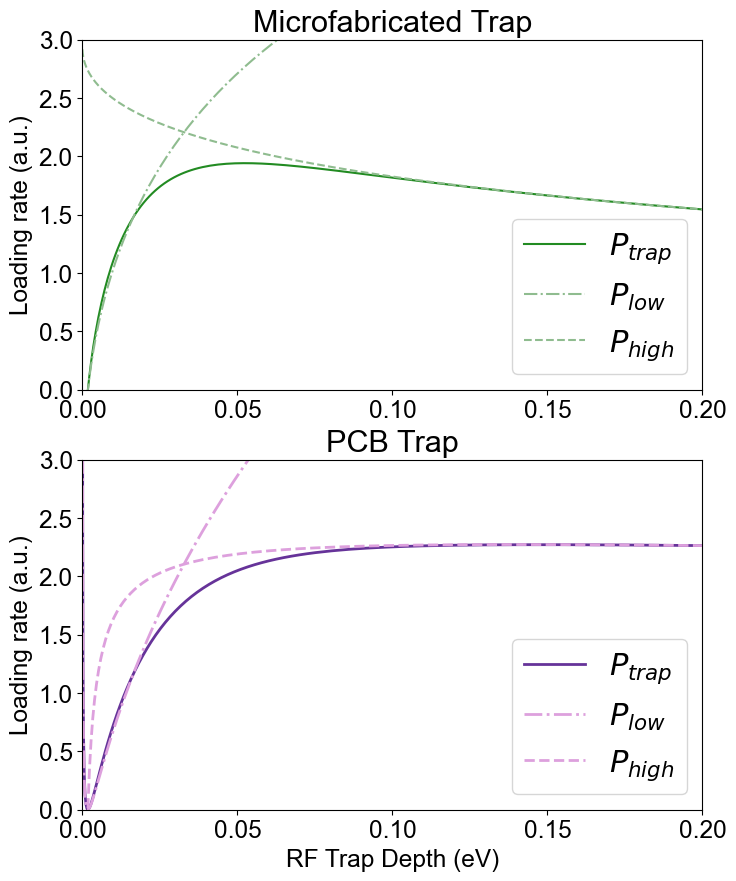}
    \caption{An analytic, harmonic estimate for the probability of loading an ion into the microfabricated trap (top) and the PCB trap (bottom) as a function of trap depth.  The low-depth limiting behavior $P_{\textrm{low}}$ (Eq. \ref{eq:p_low} and \ref{eq:p_low_pcb} for compact and large traps, respectively) is shown in dot-dashed lines, and the high-depth limiting behavior $P_{\textrm{high}}$ (Eq. \ref{eq:p_high} and \ref{eq:p_high_pcb}) is shown in dashed lines.  Both of these values are proportional to the bare trapping volume \stvbare, so this predicts that under equivalent loading conditions the PCB trap would load ions at a significantly higher rate.  The effect of the axial trapping potential is not considered in this calculation, so  $V_{\textrm{th}}=0\ $V.}
    \label{fig:prob_analytic}
\end{figure}
\begin{eqnarray}
P_{\textrm{trap}} \appropto \frac{\mathcal{V}^{(\circ)}}{\left(1+Q(\vec{x}_0)\right)^2}
 \left[\sqrt{\frac{\pi k_{\textrm{B}}T}{2mv_0^2}}\text{erf}\left(\sqrt{\frac{m}{2k_{\textrm{B}} T}}(v_{\textrm{max}}-v_0)\right)\hspace{5pt}
   \right. 
\nonumber
   \\
   \left. 
-\frac{k_{\textrm{B}}T}{2mv_{\textrm{max}}^2}\left(1-e^{-m(v_{\textrm{max}}-v_0)^2/k_{\textrm{B}}T}\right)
   \right]. ~~~~~
   \label{eq:analytic_prob_est1}
\end{eqnarray}
In relevant regions of parameter space, this is dominated by the first term
\begin{equation}
P_{\textrm{trap}} \appropto \frac{\mathcal{V}^{(\circ)}}{\left(1+Q(\vec{x}_0)\right)^2}
 \left[\sqrt{\frac{\pi k_{\textrm{B}}T}{2mv_0^2}}\text{erf}\left(\sqrt{\frac{m}{2k_{\textrm{B}} T}}(v_{\textrm{max}}-v_0)\right)\right].
 \label{eq:analytic_prob_est2}
 \end{equation}
 We estimate an optimal value for this by expanding the low-depth behavior to lowest order in $v_{\textrm{max}}-v_0$
\begin{equation}
P_{\textrm{low}}^{(\textrm{MF})} \sim \frac{v_{\textrm{max}}-v_0}{v_0\left(1+Q(\vec{x_0})\right)^2}
\label{eq:p_low}
\end{equation}
and by considering the limiting behavior of the error function at high trap depth $v_{\textrm{max}}>>v_0$, 
\begin{equation}
P_{\textrm{high}}^{(\textrm{MF})}\sim\sqrt{\frac{\pi k_{\textrm{B}}T}{2mv_0^2}}\frac{1}{\left(1+Q(\vec{x}_0)\right)^2}
\label{eq:p_high}
\end{equation}
The crossover point between these two regimes gives a rough estimate of the optimal $v_{\textrm{max}}$ and thus an optimal trap depth
\begin{equation}
E_{\textrm{opt}}^{(\textrm{MF})}\approx \frac{1}{2}m\left(v_0+\sqrt{\frac{k_{\textrm{B}} T}{m}}\right)^2
\end{equation}
For a center-of-mass velocity of $40\ \si{\meter\per\second}$ and an ablated barium temperature of 1500 K, this gives an optimal trap depth of 0.08 eV.  For a center-of-mass velocity of $40\ \si{\meter\per\second}$ and an ablated barium temperature of 50 K, this gives an optimal trap depth of 0.01 eV.  The appropriate temperature to use is discussed in Appendix \ref{sec:error}. 
 This optimal point does not depend on the trap geometry; however, the rate with which this quantity drops off at high trap depth $dP_{\textrm{high}}/dv_{\textrm{max}}$ goes as $1/Q$. Traps with larger $Q$ (which are generally physically smaller) demonstrate a faster roll-off, so it is more critical to operate close to the optimal loading point in these traps.

In the case where the PI beams are small compared to the bare trapping region (i.e. in the case of physically large traps, generally) the trapping volumes can instead be considered approximately one-dimensional.  In this case, the kinetic-energy truncated trap volume instead goes as
\begin{equation}
\mathcal{V}^{(\textrm{KE})}\approx\mathcal{V}^{(\circ)}\left(\frac{v_{\textrm{max}}-v}{v_{\textrm{max}}}\right)
\end{equation}
and the micromotion-corrected trap volume goes as
\begin{equation}
    \mathcal{V}^{(mm)}\approx\mathcal{V}^{(\textrm{KE})}\left(\frac{1}{1+Q(\vec{x}_0)}\right)
\end{equation}
so
\begin{equation}
P_{\textrm{trap}} \propto
\frac{\mathcal{V}^{(\circ)}}{1+Q(\vec{x}_0)}
\int_{0}^{v_{\textrm{max}}}\left(\frac{1}{v}-\frac{1}{v_{\textrm{max}}}\right)e^{-m(v-v_0)^2/2k_{\textrm{B}}T}dv.
\end{equation}
Again, we only consider the regime $m(v_{\textrm{max}}-v_0)^2/2<<k_{\textrm{B}}T$, so we can make the approximation 
\begin{equation}
P_{\textrm{trap}} \propto
\frac{\mathcal{V}^{(\circ)}}{1+Q(\vec{x}_0)}
\int_{0}^{v_{\textrm{max}}}\left(\frac{1}{v_0}-\frac{1}{v_{\textrm{max}}}\right)e^{-m(v-v_0)^2/2k_{\textrm{B}}T}dv
\end{equation}
which evaluates to
\begin{align*}
P_{\textrm{trap}} \propto \frac{\mathcal{V}^{(\circ)}}{1+Q(\vec{x}_0)}
 \bigg[\sqrt{\frac{\pi k_{\textrm{B}}T}{2m}}&\left(\frac{1}{v_0}-\frac{1}{v_{\textrm{max}}}\right)\times\\
 &\text{erf}\left(\sqrt{\frac{m}{2k_{\textrm{B}} T}}(v_{\textrm{max}}-v_0)\right)\bigg].
 \label{eq:analytic_prob_est3}
 \end{align*}
 At low trap depths, this expression goes as
\begin{equation}
P_{\textrm{low}}^{(\textrm{PCB})} \sim \frac{v_{\textrm{max}}-v_0}{1+Q(\vec{x}_0)}\left(\frac{1}{v_0}-\frac{1}{v_{\textrm{max}}}\right)
\label{eq:p_low_pcb}
\end{equation}
while at high depth, this goes as
\begin{equation}
P_{\textrm{high}}^{(\textrm{PCB})}\sim\sqrt{\frac{\pi k_{\textrm{B}}T}{2m}}\left(\frac{1}{v_0}-\frac{1}{v_{\textrm{max}}}\right)\frac{1}{\left(1+Q(\vec{x}_0)\right)}.
\label{eq:p_high_pcb}
\end{equation}
Recalling $Q(\vec{x}_0)$ is proportional to $\sqrt{E_{\textrm{max}}}$, $P_{\textrm{high}}^{(\textrm{PCB})}$ does not decrease as quickly with $E_{\textrm{max}}$ as $P_{\textrm{high}}^{(\textrm{MF})}$ over the relevant region of parameter space ($Q(\vec{x}_0)<1$). Further, we observe that since $P_{\textrm{high}}^{(\textrm{PCB})}$ is non-monotonic in $E_{\textrm{max}}$, the maximum of $P_{\textrm{trap}}^{(\textrm{PCB})}$ is close to the maximum of $P_{\textrm{high}}^{(\textrm{PCB})}$ at
\begin{equation}
E_{\textrm{opt}}^{(\textrm{PCB})}\approx\frac{1}{2}mv_0^2\left(1+\sqrt{1+\frac{1}{Q_0(\vec{x}_0)}}\right)^2
\end{equation}
where $Q_0(\vec{x}_0)$ is the on-axis value of Eq. \ref{eq:matheiuq} at a trap depth of $E_{\textrm{max}}=mv_0^2/2$.  The predicted value of this optimal depth for the PCB trap and a center-of-mass velocity of $50\ \si{\meter\per\second}$ is 0.31~eV, which corresponds to an rf amplitude $V_{\textrm{rf}}=580\ \si{\volt}$ at our rf drive frequency, higher than we can reach with our experimental hardware.  This is consistent with the fact that we never experimentally observed a reduction in trapping efficiency with increased rf amplitude in the PCB trap.
See Fig. \ref{fig:prob_analytic} for the expected behavior of the ion loading rate in the two traps based on the analytic estimates outlined in this Appendix.

\section{Model Parameters and Errors}
\label{sec:error}

\begin{table*}[btp]
    \centering
    \begin{tabular}{|l|r|r|r|}\hline
         & & PCB Trap & Microfabricated Trap\\
         \hline
         \multirow{4}{*}{Geometric parameters} &
            $a$ & $840~\si{\micro\meter}$ & $34~\si{\micro\meter}$\\
            & $b$ & $840~\si{\micro\meter}$ & $127~\si{\micro\meter}$\\
            & $z_0$ & $730~\si{\micro\meter}$ & $50~\si{\micro\meter}$\\
            & $z_{\textrm{esc}}$ & $1300~\si{\micro\meter}$ & $100~\si{\micro\meter}$\\ \hline
        \multirow{3}{*}{Trapping parameters} &
            $V_{\textrm{rf}}$ & $100-1000~\si{\volt}$ & $40-200~\si{\volt}$\\
            & $\Omega_{\textrm{rf}}$ & $2\pi\times7~\si{\mega\hertz}$ & $2\pi\times40~\si{\micro\meter}$ \\
            & $\omega_{\textrm{ax}}$ & $2\pi\times100~\si{\kilo\hertz}$ & $2\pi\times500~\si{\kilo\hertz}$ \\ \hline
        \multirow{4}{*}{Photo-ionization parameters} &
            $w_0$ & \multicolumn{2}{|c|}{$50~\si{\micro\meter}$} \\
            & $\gamma_1$ (Ba) & \multicolumn{2}{|c|}{$2\pi\times 18.9~\si{\mega\hertz}$~\cite{Niggli1987, Bizzarri1990}} \\
            & $\gamma_1$ (Sr) & \multicolumn{2}{|c|}{$2\pi\times32.0~\si{\mega\hertz}$\cite{Xu2003}}\\
            & $I_2$ & \multicolumn{2}{|c|}{$\sim6~\textrm{Wcm}^{-2}$} 
            \\ \hline
        \multirow{5}{*}{Source parameters} &
            $w_{\textrm{a}}$ & \multicolumn{2}{|c|}{$1~\si{\milli\meter}$} \\
            & $T$ (Ba) & \multicolumn{2}{|c|}{$1500~\si{\kelvin}$} \\
            & $v_0$ (Ba) & \multicolumn{2}{|c|}{$40~\si{\meter\per\second}$} \\
            & $T$ (Sr) & \multicolumn{2}{|c|}{$225~\si{\kelvin}$} \\
            & $v_0$ (Sr) & \multicolumn{2}{|c|}{$70~\si{\meter\per\second}$}\\ \hline
    \end{tabular}
       \caption{Parameters used in the loading model.}
    \label{tab:parameters_model}
\end{table*}

\begin{table*}[btp]
\begin{tabular}{|l|c|c|c|}
\hline
Error source & Estimated range of values & $\Delta P_{\textrm{trap}}$  \\ \hline
Stray electric field & 
$<10\ \si{\volt\per\centi\meter}$ (*),
$<1~\si{\volt\per\centi\meter}$ ($\dag$)
& 10\% to 1000\%   \\
Drive rf amplitude & 1\%
& 1\% to 10\%   \\
PI intensity & 10\%   & 10\%  \\
Source temperature & $20\%$  & $15\%$ to $20\%$ \\
Target efficiency &
$ 10\% $
& $10\%$ \\ 
\hline
\end{tabular}
\caption{Estimate of contributions to error in the model-predicted probability of loading, shown as the shaded error regions in Fig. \ref{fig:load_vs_depth_expt}.  The effects due to PI laser intensity and target efficiency are linear; all other sources contribute error that varies with trap depth within the range listed. (*) Microfabricated trap. ($\dag$) PCB trap}
\label{tab:parameter_error}
\end{table*}

  The conditions considered in the theoretical model are collected in Table \ref{tab:parameters_model}.  We generally divide these into four categories: geometric parameters, set by the electrode geometry of the trap; trapping parameters, set by the rf drive and DC trapping potentials; photo-ionization parameters, set by the PI lasers, and ablation source parameters, set by the ablation laser and the target material.  In addition to the statistical error reported in the data points, we estimate an error in the theoretical model based on estimates of the uncertainties in our model parameters.  These are summarized in Table \ref{tab:parameter_error} and shown as the shaded regions in Fig. \ref{fig:load_vs_depth_expt}.

  \textbf{Geometric parameters}  The geometric parameters are set by the fabrication process and are known to high precision.  We use $a$ and $b$ as defined in Ref.~\cite{House2008}, where $a$ is the distance separating the two symmetric rf electrodes and $b$ is the width of the rf electrodes.  The ion height $z_0$ is then determined from these parameters, as derived in \cite{House2008} and reproduced above.  

  \textbf{Trapping Parameters} 
The trapping parameters are $V_{\textrm{rf}}$, the rf drive amplitude; $\Omega_{\textrm{rf}}/2\pi$, the rf drive frequency; and $f_{\textrm{ax}} = \omega_{\textrm{ax}}/2\pi$, the axial secular frequency.  We measure the secular frequencies either by modulating the rf drive frequency at the secular frequency~\cite{Ibaraki2011} 
or by taking a Fourier transform of photon arrival times~\cite{Fan2021}.  We then extract the rf amplitude from these measurements of the radial and axial secular frequencies.  Statistical error in these frequency measurements is incorporated as horizontal error bars in Fig. \ref{fig:load_vs_depth_expt}.

Ion traps with small ion-electrode distance are highly sensitive to stray electric charge.  Typically, stray charge introduces a linear stray electric field, which can be compensated by introducing an equal and opposite field.  However, these stray fields can vary shot-to-shot over the course of an experiment.   Based on measurements before and after the experiment, we bound the uncompensated stray field introduced during the loading attempts to $E_{s}<10~\si{\volt\per\centi\meter}$ in the microfabricated trap.  In the PCB trap, we bound our uncompensated stray field by the stricter limit $E_{s}<1~\si{\volt\per\centi\meter}$, since the ion is trapped by a relatively low-frequency potential and larger stray fields would move the ion out of our detection beam.
We estimate the effect of this stray electric field on the predictive loading model by adding an additional perturbing potential
\begin{equation}
\Delta\phi_{\textrm{s}}(\vec{x}) = \vec{E}_{\textrm{s}}\cdot\vec{x}
\end{equation}
to $\phi_{\textrm{TOT}}$ and calculating the potential at the escape point of the resulting potential.  We estimate the effective trap depth with $\vec{E}_{\textrm{s}}\in\{\pm E_{\textrm{s}}\hat{x}, \pm E_{\textrm{s}}\hat{z}\}$ and incorporate the range of results into the model error, i.e. the shaded region in Fig. \ref{fig:load_vs_depth_expt}.  This effect is significant for trap depths corresponding to rf drive amplitudes near $V_{\textrm{th}}$, since a small electric field may increase or decrease the trap depth enough to enable or preclude loading ions.  At high trap depth, this effect is small. 

\textbf{Photo-ionization Parameters}
Since the first-stage PI transition is highly saturated, the dependence of the loading rate on the intensity of this laser is suppressed.  However, since the probability of trapping is linear in the intensity of the second-stage PI laser, the error is also linear in this intensity.  We include a factor of up to 10\% error due to day-to-day fluctuations in second-stage PI laser intensity.

\textbf{Source parameters}
We model the ablation plume as a thermal source with temperature $T$ and center-of-mass velocity $v_0$.  We estimate the center-of-mass velocity from the time-of-flight measurements shown in Fig.~\ref{fig:timeofflight}.  Since slowly moving atoms scatter photons multiple times while passing through the beam, we weight the data inversely by the expected number of scattering events an atom would experience passing through a $50\ \si{\micro\meter}$ radius beam.  Based on our measured thermal distribution, we estimate the temperature of the atoms to be approximately 50 K.  Prior work on barium ablation suggests, however, that the true distribution is bimodal, and that the colder component has a temperature of $1500\pm300\ \si{\kelvin}$~\cite{Rossa2009}.  For the strontium source, we include the center-of-mass velocity of the pre-cooled source, similar to what has been previously reported\cite{Sage2012, Bruzewicz2016}.  We estimate the temperature to be approximately one third of the measured temperature of a thermocouple mounted on the strontium oven, i.e. about 225 K, since the magneto-optical trap is cooled in two (out of three) directions.

A significant unknown in this experiment is the shot-to-shot surface morphology of the Ba ablation target.  Using a similar target, we observed a decay in the  $554\ \si{\nano\meter}$ fluorescence of the ablated plume after thousands of ablation pulses.  We thus include in the model the possibility of an overall linear error consistent with the expected decay of the target efficiency over the total number of ablation laser shots used in each measurement.  We also randomized the order of rf amplitudes used in loading rate measurements to avoid systematic efficiency decays due to target lifetime.


\end{document}